\documentclass[%
 aip,
 numerical,
 jmp,%
 amsmath,amssymb,
 floatfix,
reprint,
]{revtex4-1}
\usepackage{amsmath,mathtools}
\makeatletter
\renewcommand*\env@matrix[1][\arraystretch]{%
  \edef\arraystretch{#1}%
  \hskip -\arraycolsep
  \let\@ifnextchar\new@ifnextchar
  \array{*\c@MaxMatrixCols c}}
\makeatother
\usepackage{accents}

\usepackage{pgfplots}
\usepackage{epstopdf}
\usepackage{bibentry,natbib}
\usepackage{sublabel}
\usepackage{subfigure}
\usepackage{accents,amsfonts}
\usetikzlibrary{patterns}
\newcommand{\vv}[1]{{\bf #1}}
\begin{document}

\title{Using Machine Learning to Replicate Chaotic Attractors and Calculate Lyapunov Exponents from Data.}

\author{Jaideep Pathak}
\affiliation{Institute for Research in Electronics and Applied Physics, University of Maryland, College Park, Maryland 20742, USA}
\affiliation{Department of Physics, University of Maryland, College Park, Maryland 20742, USA.}
\author{Zhixin Lu}
\affiliation{Institute for Research in Electronics and Applied Physics, University of Maryland, College Park, Maryland 20742, USA}
\affiliation{Institute for Physical Science and Technology, University of Maryland, College Park,
Maryland 20742, USA.}
\author{Brian R. Hunt}
\affiliation{Institute for Physical Science and Technology, University of Maryland, College Park,
Maryland 20742, USA.}
\affiliation{Department of Mathematics, University of Maryland, College Park, Maryland 20742, USA.}
\author{Michelle Girvan}
\affiliation{Institute for Research in Electronics and Applied Physics, University of Maryland, College Park, Maryland 20742, USA}
\affiliation{Department of Physics, University of Maryland, College Park, Maryland 20742, USA.}
\affiliation{Institute for Physical Science and Technology, University of Maryland, College Park,
Maryland 20742, USA.}
\affiliation{London Mathematical Laboratory, 14 Buckingham Street, London WC2N 6DF, United Kingdom.}
\author{Edward Ott}
\affiliation{Institute for Research in Electronics and Applied Physics, University of Maryland, College Park, Maryland 20742, USA}
\affiliation{Department of Physics, University of Maryland, College Park, Maryland 20742, USA.}
\affiliation{Department of Electrical and Computer Engineering, University of Maryland, Maryland 20742, USA.}
\date{\today}
\begin{abstract}
We use recent advances in the machine learning area known as `reservoir computing' to formulate a method for model-free estimation from data of the Lyapunov exponents of a chaotic process. The technique uses a limited time series of measurements as input to a high-dimensional dynamical system called a `reservoir'. After the reservoir's response to the data is recorded, linear regression is used to learn a large set of parameters, called the `output weights'. The learned output weights are then used to form a modified autonomous reservoir designed to be capable of producing \textit{arbitrarily long} time series whose ergodic properties approximate those of the input signal. When successful, we say that the autonomous reservoir reproduces the attractor's `climate'. Since the reservoir equations and output weights are known, we can compute derivatives needed to determine the Lyapunov exponents of the autonomous reservoir, which we then use as estimates of the Lyapunov exponents for the original input generating system. We illustrate the effectiveness of our technique with two examples, the Lorenz system, and the Kuramoto-Sivashinsky (KS) equation. In particular, we use the Lorenz system to show that achieving climate reproduction may require tuning of the reservoir parameters. For the case of the KS equation, we note that as the system's spatial size is increased, the number of Lyapunov exponents increases, thus yielding a challenging test of our method, which we find the method successfully passes.
\end{abstract}
%
%
%
%
\maketitle
\begin{quotation}
There have been notable recent advances in machine learning that have proven useful for tasks ranging from speech recognition \cite{hintondeep,goodfellow} to playing of the game Go at a level surpassing the best humans \cite{silvergo}. In this paper, we build a machine learning model of a chaotic dynamical system using the neural computing framework known as reservoir computing~\cite{lukoreservoir}. We show that such a model can be used to deduce the most important quantifiers of the system's chaotic behavior, namely, its Lyapunov exponents, using only limited time series measurements of the system. We envision that such artificial intelligence based models could be used to accurately capture the complex dynamics of many geophysical, ecological, biological or economic systems that are often difficult to model from first principles.
\end{quotation}
\section{\label{sec:introduction}Introduction}
We consider the frequently occurring situation in which limited duration time series data from some dynamical process is available, but a first-principles-based model of how that data is produced is either unavailable or too inaccurate to be useful. Thus, if one is interested in diagnosing ergodic properties of the underlying processes producing the data, one is restricted to do so based only on the data itself. We call such a method ``model-free.'' Model-free analysis of dynamical time series is a long-standing subject of study in nonlinear dynamics~\cite{kantzNonlinear,ottCoping,abarbanelAnalysis}. Perhaps the most wide-spread approach uses delay-coordinate embedding~\cite{kantzNonlinear,ottCoping,abarbanelAnalysis,takens,sauerEmbedology,broomheadExtracting,brandstatedStrange,eckmannLiapunov,petrovControlling}. In this article, we discuss a very promising, entirely different approach to model-free analysis of dynamical time series. Our approach is based upon recent significant advances in the area known as \textit{machine learning}. In particular, we will apply a type of machine learning known as reservoir computing~\cite{lukoreservoir}, and, for definiteness, we focus on the problem of determining the Lyapunov exponents of the data-generating system. For this application, the key ability we require from machine learning is to replicate the ergodic properties of the system generating the input, and we call this replicating the ``climate.''

The rest of this article is organized as follows. Section~\ref{sec:RCN} reviews reservoir computing and its use for short-term prediction of chaotic time series. Section~\ref{sec:Lorenz} illustrates our method using the well-known Lorenz 1963 model~\cite{lorenzDeterministic}, and discusses the ability of reservoir computers to replicate the (long-term) climate. Section~\ref{sec:KS} uses our approach to evaluate the Lyapunov exponents of the Kuramoto-Sivashinsky (KS) equation~\cite{cohenNonlinear,kuramotoPersistent,sivashinskyLarge} with periodic boundary conditions. This system provides an example of extensive spatiotemporal chaos~\cite{crossPattern,liviDistribution,egolfRelation,pikovskyDynamic}, for which the attractor dimension and number of positive Lyapunov exponents increases linearly with the periodicity length $L$. In particular, Sec.~\ref{sec:KS} considers cases with many positive Lyapunov exponents. The paper concludes with further discussion in Sec.~\ref{sec:conclusion}.

The main conclusion of this paper is that our machine learning approach offers a very attractive model-free method for obtaining Lyapunov exponents from data. Particularly notable are our results from Sec. \ref{sec:KS} where we obtain excellent agreement for all of the positive Lyapunov exponents and many of the negative exponents for a moderately high-dimensional system. In comparison with delay coordinate embedding, we remark that our method appears to be simpler to implement, and does not appear to suffer from the problem of yielding spurious positive Lyapunov exponents (E.g., see [\onlinecite{kantz2013problem}], [\onlinecite{sauer1998spurious}]. These papers and references therein discuss the mechanism responsible for spurious positive Lyapunov exponents in delay coordinate embedding and how to fix the problem. Since this mechanism is inherently absent in our method, we do not expect, and indeed, have not found spurious positive exponents). More broadly, our paper suggests that machine learning is useful for analysis of data from chaotic systems (e.g., previous work has treated model-free machine learning for prediction of future evolution of the states of a dynamical system \cite{jaegerHarnessing} and for inference of unmeasured dynamical variables \cite{lu2017reservoir}).

\section{\label{sec:RCN}Reservoir Computers, Short Term Prediction and Attractor Climate}

\begin{figure}[htbp]
	\includegraphics[scale=.4]{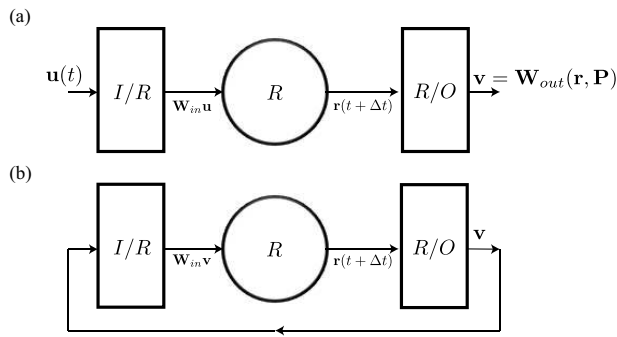}
	\caption{(a) Configuration of the reservoir in the training phase corresponding to Eqs.~\ref{eqn:1} and \ref{eqn:2}. (b) Reservoir configuration in the prediction phase corresponding to Eq.~\ref{eq:res_pred_gen}.}
\label{fig:1}
\end{figure}
Reservoir computers~\cite{lukoreservoir} originate from an idea independently put forth about 16 years ago in two papers~\cite{jaegerEcho,maassReal}. The general approach is illustrated in Fig.~\ref{fig:1}(a), which shows an input vector ${\bf u}(t)$ fed into a ``reservoir'' (labeled R in Fig.~\ref{fig:1}(a)) through an input-to-reservoir coupler (labeled I/R), with an output vector ${\bf v}$ coupled from the reservoir through an output coupler (labeled R/O). We regard the couplers as acting instantaneously and without memory (i.e., their output depends solely on their current input). Importantly, the reservoir has memory (i.e., it has internal dynamics so its state depends on its history). We assume that it receives input at discrete times $t$, and that its input $\vv{W}_{in}\vv{u}(t)$ is combined with the reservoir state $\vv{r}(t)$ to produce its output $\vv{r}(t+\Delta t)$. In general, the reservoir can be any complex dynamical system with many state variables, but here we follow Refs.~[\onlinecite{jaegerEcho},\onlinecite{maassReal}] and consider the reservoir to be a large random network with $D_r$ nodes and an $D_r\times D_r$ adjacency matrix ${\bf A}$. Specifically, we will henceforth consider the particular implementation (similar to Ref.~[\onlinecite{jaegerHarnessing}]) of Fig.~\ref{fig:1}(a) given by 
\begin{equation}
	{\bf r}(t+\Delta t) = \tanh[{\bf Ar}(t) + {\bf W}_{in}{\bf u}(t)],
	\label{eqn:1}
\end{equation}
\begin{equation}
	{\bf v}(t + \Delta t) = {\bf W}_{out}({\bf r}(t + \Delta t),{\bf P}),
	\label{eqn:2}
\end{equation}
where ${\bf r}(t)$ represents the scalar states $r_i(t)$ of the $D_r$ network reservoir nodes, ${\bf r} = [r_1,r_2,...,r_{D_r}]^T$; in Eq.~(\ref{eqn:1}), ${\bf W}_{in}$ is a $D_r \times D$ matrix, where $D$ is the dimension of ${\bf u}$; also, in Eq. (\ref{eqn:1}), for a vector $\vv{q} = (q_1, q_2, \dots )^T$ the quantity $\tanh(\vv{q})$ is the vector $(\tanh(q_1), \tanh(q_2), \dots )^T$. In Eq.~(\ref{eqn:2}), ${\bf W}_{out}$ maps the $D_r$ dimensional vector ${\bf r}$ to the output ${\bf v}$, which, for the situations considered in this article, has the same dimension $D$ as ${\bf u}$. In addition, we assume that ${\bf W}_{out}$ depends on a large number of adjustable parameters given by the elements of the matrix ${\bf P}$, and that ${\bf W}_{out}({\bf r},{\bf P})$ depends linearly on ${\bf P}$ (e.g., in past work the choice $\vv{W}_{out}(\vv{r}, \vv{P}) = \vv{P}\vv{r}$ has often been used). 

In general, the goal of the system in Fig.~\ref{fig:1}(a) is for the outputs $\vv{v}(t)$ to approximate the desired outputs, ${\bf v}_d(t)$, appropriate to the inputs ${\bf u}(t)$ (e.g., in a pattern recognition task ${\bf u}(t)$ might represent a sequence of patterns, and ${\bf v}_d(t)$ would represent classifications the patterns). To this end, during a training period, $-T\leq t\leq 0$, an input ${\bf u}(t)$ is fed into the reservoir and the resulting reservoir state evolution ${\bf r}(t)$, along with ${\bf u}(t)$, are recorded and stored as ``training data.'' Then the parameters ${\bf P}$ are chosen (``trained'') so as to approximately minimize the mean squared difference between ${\bf v}(t)$ and its desired value ${\bf v}_d(t)$. As is common in reservoir computing, we use the Tikhonov regularized regression procedure \cite{tikhonov1977solutions} to find an output matrix $\vv{P}$, that minimizes the following function,
\begin{equation}
\sum_{-T\leq t\leq 0}\vert\vert{\bf W}_{out}(\vv{r}(t), \vv{P}) - {\bf v}_d(t)\vert\vert^2 + \beta \lVert \vv{P} \rVert^2,
	\label{eq:minimization}
\end{equation}
where $\lVert \vv{P} \rVert^2$ denotes the sum of the squares of elements of $\vv{P}$. The regularization constant $\beta>0$ discourages overfitting by penalizing large values of the fitting parameters (In Sec.~\ref{sec:KS} we used a value $\beta > 0$, but for Sec.~\ref{sec:Lorenz} we found that using $\beta = 0$ was sufficient). For a given task, one hopes that for large enough $D_r$ and $T$, the system in Fig.~\ref{fig:1}(a) will yield subsequent ($t>0$) outputs ${\bf v}(t)$ that closely approximate the desired ${\bf v}_d(t)$. Because ${\bf W}_{out}({\bf r},{\bf P})$ is taken to be linear in ${\bf P}$, the problem of determining the parameters ${\bf P}$ that minimize Eq.~(\ref{eq:minimization}) is one of linear regression for which there are well-established techniques \cite{yan2009linear}. This approach has been shown to work extremely well for a wide variety of tasks~\cite{lukoreservoir}.

We now consider the task of prediction for the case where ${\bf u}(t)$ depends on the state of some deterministic dynamical system. This problem was originally considered in the reservoir computer framework by Jaeger and Haas~\cite{jaegerHarnessing}. The idea is to take the desired output to be the same as the input, ${\bf v}_d(t + \Delta t) = {\bf u}(t + \Delta t)$. When one wishes to commence prediction at $t = 0$, the configuration is switched from that in Fig.~\ref{fig:1}(a) to that in Fig.~\ref{fig:1}(b), and the reservoir system is run autonomously according to the following equation.
\begin{align}\label{eq:res_pred_gen}
\vv{r}(t+\Delta t) &= \tanh \left[\vv{A}\vv{r}(t) + \vv{W}_{in}\vv{W}_{out}(\vv{r}(t), \vv{P}) \right].
\end{align}
The output of the autonomous reservoir, $\vv{v}(t) = \vv{W}_{out}(\vv{r}(t), \vv{P})$, gives the predicted value ${\bf u}(t)$ for $t > 0$.  Jaeger and Haas~\cite{jaegerHarnessing}, using the example of the Lorenz system~\cite{lorenzDeterministic}, indeed verified that this prediction scheme works and gives good short term predictions. As expected, the chaotic amplification of small errors leads to eventual breakdown of the prediction, limiting the prediction time. However, as shown in the next two sections, following this breakdown of short-term prediction, the evolution of ${\bf v}(t)$ often provides an accurate approximation for the climate corresponding to ${\bf u}(t)$, and can be used in particular to compute Lyapunov exponents of the process that generated $\vv{u}(t)$.

\section{\label{sec:Lorenz}Example 1: the Lorenz System and the Question of Whether the Climate is Replicated}
In this section we illustrate the capability of our technique to replicate the ``climate'' of the Lorenz 1963 system~\cite{lorenzDeterministic},
\begin{equation}
\begin{aligned}
	\dot{x} &= 10(y-x),\\
	\dot{y} &= x(28-z)-y,\\
	\dot{z} &= xy-8z/3.
\end{aligned}
\label{eqn:Lorenz}
\end{equation}

We construct and train reservoir computers with input ${\bf u}=(x,y,z)^T\in \mathbb{R}^3$ and output ${\bf v}\in\mathbb{R}^3$, following Sec.~\ref{sec:RCN}. The reservoir network is built from a sparse random Erd\H{o}s-R{\'e}nyi network whose average degree is $ d =6$. Each non-zero element in the adjacency matrix is drawn independently and uniformly from $[-a,a]$, and $a>0$ is adjusted so that the spectral radius of ${\bf A}$ (the largest magnitude of its eigenvalues) has a desired value $\rho$. During the training phase, $-T\leq t\leq0$ (where $T=100$), the reservoir computer evolves following Eq.~(\ref{eqn:1}) with $\Delta t = 0.02$. In this Lorenz example, the reservoir output ${\bf v}(t) = {\bf W}_{out}({\bf r}(t),{\bf P})$ is defined as
\begin{align}
\vv{v}(t) =
\begin{bmatrix}
v_1(t)\\ v_2(t) \\v_3(t) 
\end{bmatrix}
=
\begin{bmatrix}
{\bf p}_1  {\bf r}(t)\\
{\bf p}_2 {\bf r}(t)\\
{\bf p}_3 \tilde{{\bf r}}(t)
\end{bmatrix}
\label{eqn:output}
\end{align}
where ${\bf p}_1$, ${\bf p}_2$, and ${\bf p}_3$ are the rows of the $3 \times D_r$ matrix $\vv{P}$. The quantity $\tilde{{\bf r}}$ in the third line of Eq.~(\ref{eqn:output}) is defined in a way such that the first half of its elements are the same as that of ${\bf r}$, i.e., $\tilde{r}_i=r_i$ for half ($D_r/2$) of the reservoir nodes, while $\tilde{r}_i=r_i^2$ for the remaining half of the reservoir node (Our use here of $\tilde{\vv{r}}(t)$, rather than $\vv{r}(t)$, to predict $z(t)$ is related to the $x \rightarrow -x$, $y \rightarrow -y$ symmetry of the Lorenz equations as discussed in Ref.~[\onlinecite{lu2017reservoir}]).  

After we compute $\vv{r}(t)$ for the training period, $-T \leq t \leq 0$, we calculate the output weight parameters ${\bf P}$ that minimize the function in Eq.~(\ref{eq:minimization}) with the desired output being the state variables from the Lorenz system, ${\bf v}_d(t)=[x(t),y(t),z(t)]^T$ (in an actual physical experiment, we assume $\vv{u}(t) = \vv{v}_d(t)$ to have been measured for $-T \leq t\leq 0$). After we find the output weights, we evolve the reservoir with the reconfigured reservoir system (Fig.~\ref{fig:1}(b)).

\begin{table}
\begin{tabular}{c|c||c|c}
Parameter & Value & Parameter & Value\\
\hline
$D_r$ & 300 & $d$ & 6\\
$T$ & 100 &$\Delta t$ & 0.02\\
$T/\Delta t$ & 5000  & $\beta$ & 0 \\
$\rho$ & 1.2 & $\sigma$ & 0.1
\end{tabular}
\caption{Standard reservoir parameters used for a successful climate replication of the Lorenz system (referred to in the text as the $R1$ reservoir). The $R2$ reservoir uses the same parameters with a different spectral radius, $\rho = 1.45$. }
\label{tab:reslrz} 
\end{table}

\begin{figure}[htbp]
	\includegraphics[scale=.295]{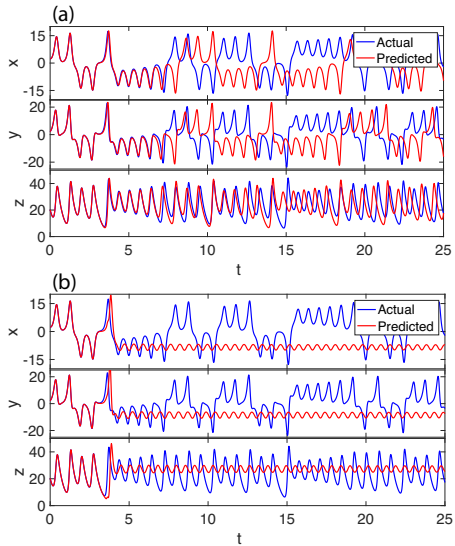}
	\caption{(a) The state prediction (red) of the R1 reservoir and the actual trajectories (blue) of the Lorenz system for $0<t\leq 25$. The spectral radius of the reservoir is $1.2$. (b) The state prediction (red) of the R2 reservoir and the actual trajectories (blue) of the Lorenz system for $0<t\leq 25$. The spectral radius of the reservoir is $1.45$.}
\label{fig:lrz_climate}
\end{figure}

Following the above described procedure, We now report and compare results for two simulations using reservoir configurations with $\rho=1.2$ (denoted R1) and $\rho=1.45$ (denoted R2). The prediction for $0<t\leq 25$ for both trained reservoirs are shown in Fig.~\ref{fig:lrz_climate}(a) (R1 with $\rho=1.2$) and Fig.~\ref{fig:lrz_climate}(b) (R2 with $\rho=1.45$). Both reservoirs R1 and R2 generate correct short-term predictions and then deviate from the actual Lorenz trajectories, which is expected since any small error grows exponentially due to the chaotic dynamics of the Lorenz system. However, after the failure of the short-term prediction, the two reservoirs show qualitatively different dynamical patterns. In Fig.~\ref{fig:lrz_climate}(a), it seems that, after $t\approx 7$, although the R1 prediction deviates from the actual trajectory, the long-term dynamics appears to resemble that of the original Lorenz system. In contrast, as shown by Fig.~\ref{fig:lrz_climate}(b), this is clearly not the case for R2.

\begin{figure}[htbp]
	\centering
\includegraphics[width = 0.5\textwidth]{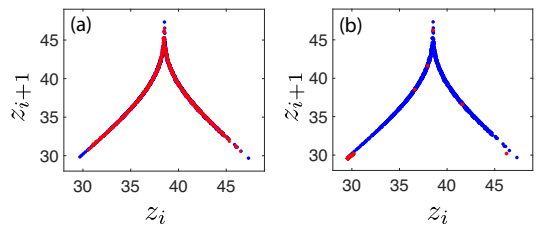}
	\caption{The return map of the actual and the predicted $z$-coordinate of the Lorenz system. This plot is made with time series of length $1000$, where the blue dots are from the actual Lorenz system, and the red dots overlaying the blue dots are from the prediction. The left panel shows the return map of the long term prediction of the R1 reservoir with $\rho=1.2$, while the right panel is from the R2 reservoir with $\rho=1.45$.}
\label{fig:returnMap}
\end{figure}

In Fig.~\ref{fig:returnMap} we present a more accurate test than visual inspection of Figs.~\ref{fig:lrz_climate}(a) and \ref{fig:lrz_climate}(b) for correctness of the climate. To do this, we follow Lorenz's procedure of plotting the return map of successive maxima of $z(t)$. We first obtain $z(t)$ for a long period of time, $0<t<1000$, for both the actual and the predicted time series. We then locate all local maxima of the actual and predicted $z(t)$ in time order and denote them $[z_1,z_2,...,z_m]$. Then, we plot consecutive pairs of those maxima $[z_i,z_{i+1}]$ for $i=1,...,m-1$ as dots in Figs.~\ref{fig:returnMap}. The blue dots in both panels of Figs.~\ref{fig:returnMap} are from the actual Lorenz system, while the red dots \textit{printed over} the blue dots are from the reservoir output prediction ($v_3$) of $z(t)$. As confirmed by Fig.~\ref{fig:returnMap}(a) the red dots produced by the R1 reservoir continue to fall on top of the blue dots (from the actual Lorenz system) throughout the entire run time ($0<t<1000$). In contrast, Fig.~\ref{fig:returnMap}(b) shows that the blue dots remain largely uncovered, because, as indicated in the third panel of Fig.~\ref{fig:lrz_climate}(b), the maximum value of $z(t)$ for $t>5$ is at a fixed point $z_{max}\approx 30$. Thus the R1 reservoir very accurately succeeds in reproducing the long-time climate of the attractor, while the R2 reservoir does not, and this is so even though both setups are apparently capable of producing useful short term predictions. (We have also obtained similar results for many other simulations.) Thus some parameter adjustment may be necessary to avoid unsuccessful reproduction of the climate. Fortunately, we usually find that when the climate is not reproduced it is fairly evident (as in Fig.~\ref{fig:lrz_climate}(b), as well as Fig.~\ref{fig:bad_climate} of the next section). More quantitatively, a promising general means of assessing whether the reservoir system has succeeded in mimicking the climate is to first use the training data to obtain finite-time estimates of the system's ergodic properties (e.g., frequency-power spectra, time correlations, moments etc.). Once this is done, one can test whether those estimates are consistent with determinations of the same quantities obtained from the long-term reservoir dynamics. Section \ref{sec:KS} provides such an assessment for the Kuramoto-Sivashinsky system.
\begin{table}
\begin{tabular}{c|c|c|c}
&Actual Lorenz System&R1 System&R2 System\\
\hline
$\Lambda_1$ & $0.91$ & $0.90$ & $0.01$ \\
$\Lambda_2$ & $0.00$ & $0.00$ & $-0.1$ \\
$\Lambda_3$ & $-14.6$ & $-10.5$ & $-9.9$ \\
\end{tabular}
\caption{Three largest Lyapunov exponents $\Lambda_1\geq\Lambda_2\geq\Lambda_3$ for the Lorenz system (Eq.~(\ref{eqn:Lorenz})), and for the reservoir set up in the configuration of Fig.~\ref{fig:1}(b) for R1 and R2. Since the reservoir system that we employ is a discrete time system, while the Lorenz system is a continuous system, for the purpose of comparison, $\Lambda_1$, $\Lambda_2$, and $\Lambda_3$ are taken to be per unit time; that is, their reservoir values (columns 2 and 3) are equal to the reservoir Lyapunov exponents calculated on a per iterate basis divided by $\Delta t$.}
\label{tab:LorenzLYP}
\end{table}

The reservoir in the autonomous configuration of Fig.~\ref{fig:1}(b) represents a known discrete-time, $D_r$-dimensional dynamical system (since we know ${\bf W}_{in}$, ${\bf A}$, and the output parameters $\vv{P}$ determined by the training). We compute the equations for the evolution of the tangent map corresponding to Eq.~(\ref{eq:res_pred_gen}) and evolve a set of $m$ mutually orthogonal tangent vectors $\vv{R}(t) = \lbrace \delta \vv{r}_j \rbrace_{j = 1}^m$ along with Eq.~(\ref{eq:res_pred_gen}). We then compute the largest $m$ Lyapunov exponents of the reservoir dynamical system in the the configuration shown in Fig. \ref{fig:1}(b) using a standard algorithm based on $QR$ decomposition (e.g., see Ref.~[\onlinecite{abarbanelAnalysis}]) of the matrix $\vv{R}(t)$. The two right-most columns of Table~\ref{tab:LorenzLYP} show the three largest Lyapunov exponents, $\Lambda_1\geq\Lambda_2\geq\Lambda_3$, of the reservoir system in the autonomous configuration, Fig.~\ref{fig:1}(b), for the R1 reservoir (for which climate reproduction succeeds), and for the R2 reservoir (for which climate reproduction fails). 

Comparing the Lyapunov exponents of the Lorenz system (first column of Table~\ref{tab:LorenzLYP}) with those of the R1 reservoir, we see that the largest Lyapunov exponent of the R1 reservoir is a good approximation to the largest Lyapunov exponent of the Lorenz system. Also, consistent with the small value of $\Delta t$, the reservoir dynamics approximates that of a flow for which $\Lambda_2$ should be (and is) approximately zero. On the other hand, we see that the third Lyapunov exponent of the R1 system is less negative than the negative Lyapunov exponent of the true Lorenz system. In contrast with the good agreement of the $\Lambda_1$ values for the Lorenz system and the R1 reservoir, the positive Lyapunov exponent of the Lorenz system fails to be reproduced by the R2 system whose largest Lyapunov exponent is approximately zero; this is consistent with the observation from Fig.~\ref{fig:lrz_climate}(b) that the long term reservoir attractor for R2 appears to be a periodic orbit. 

The significant conclusion from the above is that the R1 system, as a result of successfully reproducing the climate, can be utilized to obtain an approximation to the positive and zero Lyapunov exponents of the process generating its input. We note, however, that the R1 system does not accurately reproduce the true negative Lyapunov exponent of the Lorenz attractor. 

The inaccurate reservoir estimation of $\Lambda_3$, noted above, can be understood by noting that, although the return map in Fig.~\ref{fig:returnMap} appears to be a curve, this apparent ``curve'' must, as noted by Lorenz~\cite{lorenzDeterministic}, actually have some small width. The R1 reservoir succeeds in \textit{approximating} the attractor of the Lorenz system as reflected by its apparent good reproduction of the return map shown in Fig.~\ref{fig:returnMap}(a). In order to do this, however, the reservoir need not reproduce the very thin transverse structure within the apparent curve. Since, this very thin structure, as we next discuss, is the primary orbital evidence of the value of $\Lambda_3$, one might not expect the reservoir to accurately reproduce this very negative Lyapunov exponent. Specifically, using the Kaplan-Yorke formula for the information dimension~\cite{kaplan1979chaotic} of the fractal Lorenz attractor, we obtain a dimension of $[2+(\Lambda_1/\vert \Lambda_3 \vert)]=2.06$, corresponding to $1.06$ for the dimension of the structure in the return map (Fig.~\ref{fig:returnMap}(a)). This dimension is very close to one, in agreement with the approximate curve-like character of the return map. However, close examination of the return map ``curve'' of the Lorenz attractor has previously shown that, within its thickness, there is a fractal set of small transverse dimension (presumably $\Lambda_1/\vert\Lambda_3\vert=0.06$). On the other hand, the Kaplan-Yorke dimension for the return map for the climate of the R1 reservoir attractor is about $1.09$. Since the primary orbital difference reflected by differing values of $\Lambda_3$ is the difference in very thin structure features of the return map that have only a small effect on the climate dynamics, it is not surprising that the R1 reservoir, while giving a good approximation to the true climate of the Lorenz system, gives only a rough approximation of $\Lambda_3$.
\section{Example 2: The task of determining a large number of Lyapunov exponents of a high dimensional  spatiotemporal chaotic system from data}\label{sec:KS}
We now consider a modified version of the Kuramoto-Sivashinsky (KS) system defined by the partial differential equation for the function $y(x,t)$
\begin{align}\label{eq:ks}
y_t = -yy_x - \left[ 1+ \mu\cos\left(\frac{2\pi x}{\lambda} \right) \right]y_{xx} - y_{xxxx},
\end{align}
in the region $0 \leq x < L$ with periodic boundary conditions, $y(x,t) = y(x+L,t)$, and $\lambda$ chosen so that $L$ is an integer multiple of $\lambda$. This equation reduces to the standard KS equation when $\mu = 0$. The cosine term makes the equation spatially inhomogeneous. We will subsequently consider the cases $\mu =0$ and $\mu \neq 0$ in order to discuss the effect of the symmetries of the KS equation on the learning dynamics of the reservoir computer.

By numerically integrating Eq. (\ref{eq:ks}) on an evenly spaced one-dimensional grid of size $Q$, we obtain a discretized multivariate data set of $Q$ time series,
\begin{align}
\vv{u}(t) &= \left[y(\Delta x,t), y(2\Delta x,t), \dots , y(Q\Delta x,t) \right]^T,\\
\Delta x &= L/Q. \nonumber
\end{align}
As in the case of the Lorenz equations discussed in Sec. \ref{sec:Lorenz}, we consider the situation where we have access to the time series data but do not have information about the dynamical equation that generated the time series. In the absence of a model, we will use the data to train a reservoir computer to emulate the behavior of the true dynamical system, in this case Eq.~(\ref{eq:ks}). 

\begin{table}
\begin{tabular}{c|c||c|c}
Parameter & Value & Parameter & Value\\
\hline
$D_r$ & 9000 & $d$ & 3\\
$T$ & 20000 &$\Delta t$ & 0.25\\
$T/\Delta t$ & 80000  & $\beta$ & 0.0001 \\
$\rho$ & 0.4 & $\sigma$ & 0.5
\end{tabular}
\caption{Reservoir parameters used for the successful replication of the climate of the Kuramoto-Sivashinsky system shown in Fig.~\ref{fig:ksst}.}
\label{tab:resks} 
\end{table}
The reservoir network is as described in Sec. \ref{sec:RCN} with the parameters listed in Table \ref{tab:resks}. In the training phase, Fig. \ref{fig:1}(a), we evolve the reservoir according to Eq. (\ref{eqn:1}) from $t = -T$ to $t = 0$. Next, we use Tikhonov regularized regression (see Eq.~(\ref{eq:minimization})) to compute the output parameters, $\vv{P}$ such that $\vv{W}_{out}(\vv{r}, \vv{P}) =  \vv{P} \tilde{\vv{r}}(t) \simeq \vv{u}(t)$ for $-T\leq t <0$. Here $\tilde{\vv{r}}$ is a $D_r$-dimensional vector such that the $i^{\text{th}}$ component of $\tilde{\vv{r}}$ is $\tilde{r}_i = r_i$ for half the reservoir nodes and $\tilde{r}_i = r_i^2$ for the remaining half. With the output parameters determined, we let the reservoir evolve autonomously for $t>0$ as shown in Fig. \ref{fig:1}(b) according to Eq.~(\ref{eq:res_pred_gen}). 

\begin{figure}
\includegraphics[width = 0.4\textwidth]{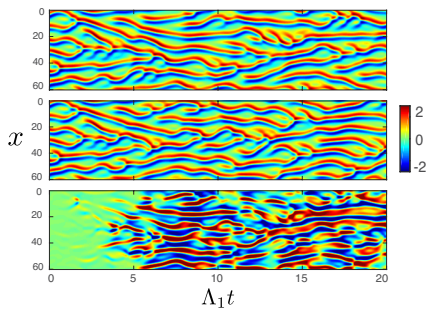}
\caption{Top panel: True state, $y(x,t)$, of the standard KS system after $t = 0$. Middle panel: Reservoir prediction. Bottom panel: Difference between the true state and the reservoir prediction. The parameters of the KS equation are $L=60$, $\mu = 0$. $\Lambda_1$ denotes the largest Lyapunov exponent.}
\label{fig:ksst}
\end{figure}

The predictions made by the reservoir system for $t > 0$ are given by, $\vv{W}_{out}(\vv{r}(t), \vv{P})$. Figure \ref{fig:ksst} shows the time evolution of one such reservoir prediction for $t > 0$ (middle panel), along with the true state (top panel) of the KS equation and the deviation (bottom panel) of the reservoir prediction from the true state (i.e., the difference between the top panel and the middle panel) Note that in Fig.~\ref{fig:ksst} time (the horizontal axis) is in units of the Lyapunov time ($\Lambda_1^{-1}$, where $\Lambda_1$ is the largest Lyapunov exponent of the KS attractor). We see that the reservoir gives good short term prediction for about 5 multiples of the Lyapunov time. A visual inspection of Fig.~\ref{fig:ksst} suggests that the reservoir prediction may have also learned the correct `climate' of the KS system even after the state of the reservoir dynamical system has diverged from the true state of the KS system. 

\begin{figure}
\includegraphics[width = 0.4\textwidth]{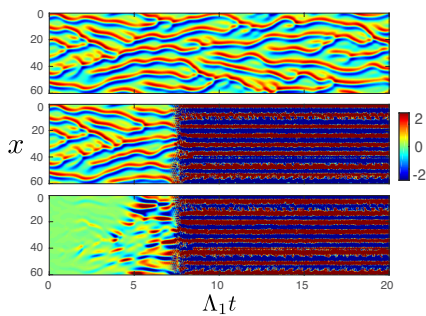}
\caption{Top panel: True state, $y(x,t)$, of the standard KS system after $t = 0$. Middle panel: Reservoir prediction with a reservoir of size $D_r = 5000$ and $\rho = 3.1$. The rest of the parameters are as given in Table \ref{tab:resks}. Bottom panel: Difference between the reservoir prediction and the true KS state. We see that in this case, the reservoir gives us an accurate short term prediction (i.e., the `weather') but the long term `climate' of the autonomous reservoir dynamical system does not resemble the climate of the true KS system for this poorly chosen set of parameters. $\Lambda_1$ denotes the largest Lyapunov exponent.}
\label{fig:bad_climate}
\end{figure}

Figure~\ref{fig:bad_climate} shows an example of an alternate scenario for another set of the reservoir parameters ($\rho = 3.1$, $D_r = 5000$ with the rest of the parameters as shown in Table~\ref{tab:resks}). In this case, the reservoir still predicts accurately for a short period of time. However, the long term climate of the signal generated by the reservoir is no longer similar to that of the true KS climate. 

\begin{figure}
\includegraphics[width = 0.4\textwidth]{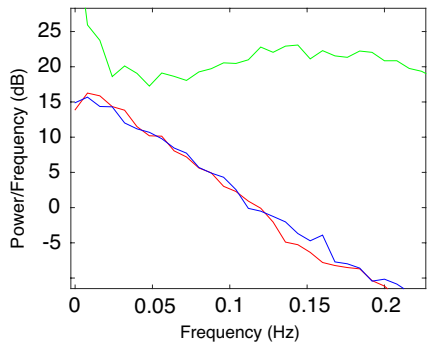}
\caption{Power spectrum of the KS training data (blue), of the reservoir prediction with the same parameters as in Fig.~\ref{fig:ksst} (red), and of the reservoir prediction with parameters as in Fig.~\ref{fig:bad_climate} (green). All power spectra have been computed at a single spatial gridpoint from a time series of length 15000 $\Delta t$ time steps. The power spectra are smoothed by dividing a time series into 30 intervals, computing the power spectrum of each interval and then averaging over all the intervals.}
\label{fig:power}
\end{figure}

A more quantitative assessment of the climate reproduction can be obtained by calculating the power spectrum of the reservoir prediction and comparing it with the power spectrum of the training data. Figure~\ref{fig:power} shows the power spectrum of the training data, along with the power spectrum of the dynamics of the autonomous reservoir system in Figs.~\ref{fig:ksst} and \ref{fig:bad_climate}. We see that the reservoir system corresponding to Fig.~\ref{fig:ksst} succeeds in reproducing the training data power spectrum, thus indicating that the long term system orbit reproduces the climate of the training data. On the other hand, the power spectrum of the reservoir system corresponding to Fig.~\ref{fig:bad_climate} confirms our visual assessment that this reservoir system fails to reproduce the climate of the training data.
 
\begin{figure*}
\centering
\includegraphics[width = 0.8\textwidth]{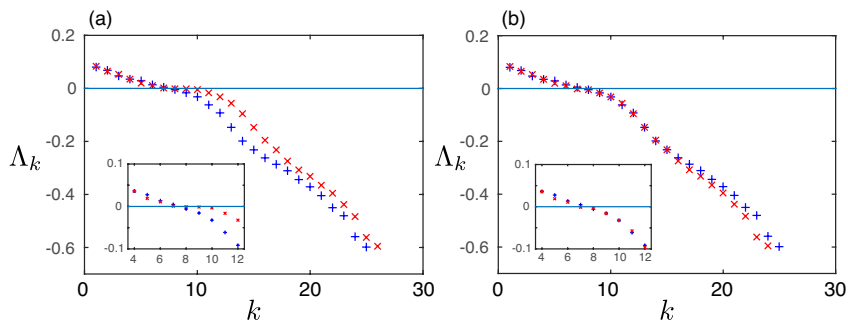}
\caption{(a) Estimating the Lyapunov exponents of the homogeneous ($\mu = 0$) KS equation. First 26 Lyapunov exponents of the trained reservoir dynamical system running in autonomous prediction mode (blue `$+$' markers) and the standard (i.e., $\mu = 0$) KS system (red `$\times$' markers). The parameters of Eq.~(\ref{eq:ks}) are $L = 60$, $\mu = 0$. (b) The same plot as (a), except, the two near-zero exponents of the KS system ($\Lambda_7$ and $\Lambda_8$) are removed from the spectrum. Inset: a close up of the spectra around the zero crossing. All Lyapunov exponents in this figure and Fig.~\ref{fig:kslyaps_p1} were computed from a trajectory of length 10000 $\Delta t$ time steps, which we found to be sufficiently long for convergence.}
\label{fig:kslyaps_p0}
\end{figure*}

\begin{figure}
\includegraphics[width = 0.4\textwidth]{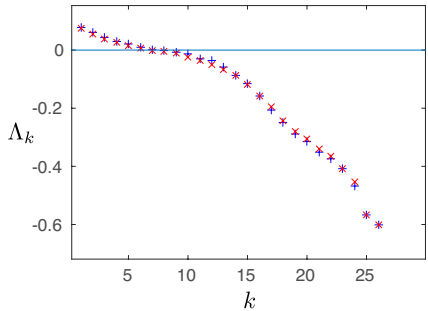}
\caption{Estimating the Lyapunov exponents of the inhomogeneous ($\mu >0$) KS equation. First 26 Lyapunov exponents of the trained reservoir dynamical system running in autonomous prediction mode (blue `$+$' markers) and the modified (i.e., $\mu > 0$) KS system (red `$\times$' markers). The parameters of Eq.~(\ref{eq:ks}) are $L = 60$, $\mu = 0.1$ and $\lambda = 15$.}
\label{fig:kslyaps_p1}
\end{figure}
Similar to what was done in Sec. \ref{sec:Lorenz}, we use our complete knowledge of the dynamics of the reservoir computer to evaluate its Lyapunov exponents. By independently evaluating the Lyapunov exponents directly from the KS equation, Eq.~(\ref{eq:ks}), we obtain the true Lyapunov exponents and compare them with the corresponding Lyapunov exponents of the reservoir dynamical system.

Figure \ref{fig:kslyaps_p0}(a) shows the Lyapunov spectrum of the standard ($\mu = 0$) KS system with $L=60$ (red `$\times$' markers), where, by definition the subscript $k$ is such that $\Lambda_k \geq \Lambda_{k+1}$. The Lyapunov exponents of the reservoir trained to emulate this system are shown on the same axes (blue `$+$' markers). We observe that the positive Lyapunov exponents of the reservoir system match the corresponding exponents of the KS system very well. However, the negative exponents of the two systems do not seem to agree with each other at first glance. We argue below that the standard KS system has three zero Lyapunov exponents, and we posit that the reservoir is unable to reproduce two of them. Indeed, Fig. \ref{fig:kslyaps_p0}(b) shows that if we remove the two of the computed exponents closest to zero ($\Lambda_7$ and $\Lambda_8$) for the KS system, the negative Lyapunov exponents of the reservoir system match those of the KS system very well.

We show now that when $\mu = 0$ (as for Fig.~\ref{fig:kslyaps_p0}), the standard KS equation (\ref{eq:ks}) has three zero Lyapunov exponents associated with three continuous symmetries, namely time-translation invariance, space-translation invariance and the so-called Gallilean invariance. Time and space translation invariance imply that if $y(x,t)$ is a solution, then so are $y(x,t+t_0)$ and $y(x+x_0,t)$. By Gallilean invariance, we mean that for every solution $y(x,t)$ of the KS equation and an arbitrary constant $v$, $y(x-vt,t) + v$ is also a solution. This can be verified by direct substitution in Eq.~(\ref{eq:ks}) with $\mu = 0$. Replacing $t_0$, $x_0$, and $v$ by differentials ($t_0 \rightarrow \delta t_0$,  $x_0 \rightarrow \delta x_0$, $v \rightarrow \delta v$), we have that, $\delta y(x,t) = \frac{\partial y(x,t)}{\partial t} \delta t_0$, $\delta y(x,t) = \frac{\partial y(x,t)}{\partial x} \delta x_0$ and $\delta y(x,t) = \left[1 - t \frac{\partial y(x,t)}{\partial x}\right] \delta v$ all represent perturbations, $y(x,t) + \delta y(x,t)$, of Eq. (\ref{eq:ks}) that are, to linear order in the differentials, solutions of Eq.~(\ref{eq:ks}). That is, all three of these $\delta y(x,t)$ are solutions of the variational equation, $\delta y_t + \delta y y_x + y \delta y_x + \delta y_{xx} + \delta y_{xxxx} = 0$. Furthermore, since the original solution $y(x,t)$ does not decay exponentially to zero, nor increase exponentially to infinity, we conclude that these three expressions for $\delta y$ represent Lyapunov vectors with zero Lyapunov exponents.

To see why the reservoir does not reproduce the Gallilean symmetry-associated zero Lyapunov exponent in the $\mu = 0$ case, notice that there is a corresponding conserved quantity $c = \int y(x,t) dx$. A particular KS system trajectory in phase space is thus restricted to a hypersurface with a constant value of $c$ (say, $c = c_0$). Since the reservoir is trained with data from a single trajectory, it does not learn the dynamics of perturbations that take the trajectory off the $c_0$ hypersurface. We are not certain why the reservoir does not reproduce both of the other two zero exponents.

As a further example that does not have additional symmetries beyond time-translation, we consider (Fig.~\ref{fig:kslyaps_p1}) a KS equation with a nonzero value of $\mu$ ($L = 60, \lambda = 15, \mu =0.1$). As before, we train the reservoir using the time series data from the symmetry broken KS equation. After training, we run the reservoir in autonomous prediction mode (Fig. \ref{fig:1}(b)) and calculate its Lyapunov spectrum. Figure \ref{fig:kslyaps_p1} shows that the reservoir reproduces the Lyapunov spectrum of the true KS system accurately in this case. Notably, in contrast with the case $\mu =0$, this good agreement is obtained without the need of discarding two zero Lyapunov exponents. We continue to use this modified KS system in the experiments described below. For the cases shown in Figs.~\ref{fig:kslyaps_p0}(b) and \ref{fig:kslyaps_p1}, the information dimension of the attractor, as computed from the Kaplan-Yorke conjecture~[\onlinecite{kaplan1979chaotic}], is about $D_{KY} \approx 15$ (roughly, the value of $k$ at which $\sum_{j=1}^k \Lambda_j$ first becomes negative). We see from Fig.~\ref{fig:kslyaps_p0}(b) and Fig.~\ref{fig:kslyaps_p1} that the reservoir continues to give reasonable estimates of $\Lambda_k$ even for $k > D_{KY}$. This was somewhat surprising to us, especially in view of the inaccurate reservoir estimate of $\Lambda_3$ in Sec.~\ref{sec:Lorenz}.

\begin{figure}
\centering
\includegraphics[width = 0.35\textwidth]{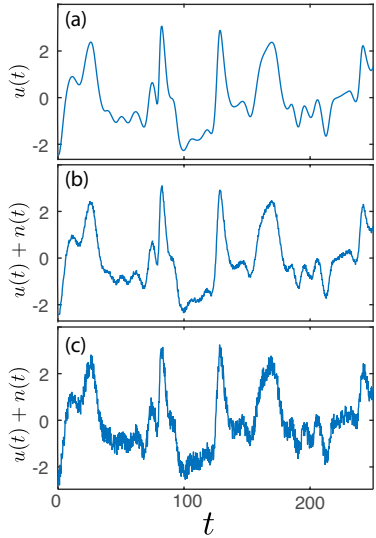}
\caption{(a) Single scalar component $u(t)$ of the time series $\vv{u}(t)$ generated from the KS system (Eq. (\ref{eq:ks})) with $L = 60$, $\lambda = 15$ and $\mu = 0.1$. The time series in (a) with added noise, $u(t) + n(t)$, of noise strengths $f=0.05$ and $f = 0.2$ are shown in (b) and (c) respectively.}
\label{fig:noisy_data}
\end{figure}

\begin{figure}
\centering
\includegraphics[width = 0.4\textwidth]{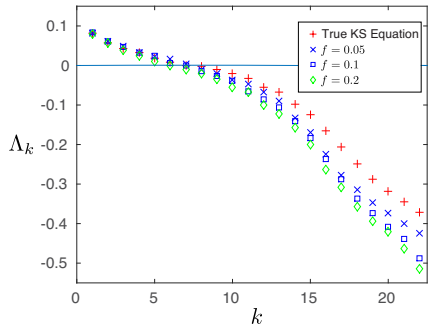}
\caption{Lyapunov exponents of the reservoir trained on noisy data from the KS system ($L = 60$, $\lambda = 15$, $\mu = 0.1$ ). The strength of the noise added to the training data is indicated in the legend.}
\label{fig:noisy_lyaps}
\end{figure}

We now consider the effect of additive measurement noise on our Lyapunov exponent calculation scheme. We simulate measurement noise by adding a random vector $\vv{n}(t)$ to the training data set $\vv{u}(t)$ for all values of $t$. That is, at every time step $\Delta
 t$, we replace $\vv{u}$ in Eq. (\ref{eqn:1}) by $\vv{u} + \vv{n}$, and we replace $\vv{v}_d = \vv{u}$ used in Eq. (\ref{eq:minimization}) by $\vv{v}_d = \vv{u} + \vv{n}$. The scalar elements $n_j(t)$ of the vector $\vv{n}(t)$, for each value of $j$ and $t$, are independent, identically distributed uniform random variables in the interval $[-\alpha, \alpha]$. The constant $\alpha$ is chosen so that the RMS value of the noise is $f$ times the RMS value of the noise-free signal $\vv{u}(t)$. Figure \ref{fig:noisy_data}(a) shows the noise-free time series at a single grid point, while Figs. \ref{fig:noisy_data}(b) and \ref{fig:noisy_data}(c) show the same time series with added noise of strength $f = 0.05$ and $f = 0.2$, respectively. We calculate the Lyapunov exponents of the reservoir as described above. Figure \ref{fig:noisy_lyaps} shows the Lyapunov spectrum when the noise level $f$ is varied from $0.05$ to $0.20$ along with the true Lyapunov spectrum of the KS equation. We see that the reservoir results for the positive Lyapunov exponents are quite robust to noise for $f \leq 0.2$, but that the negative exponents are increasingly depressed to more negative values as $f$ increases.

\begin{figure}
\centering
\includegraphics[width = 0.4\textwidth]{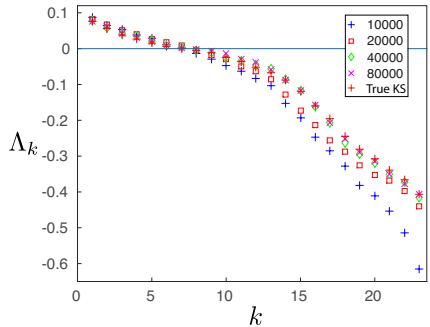}
\caption{The Lyapunov spectrum of the reservoir trained using varying lengths of training data from Eq.~(\ref{eq:ks}) with parameters $L = 60$, $\lambda = 15$ and $\mu=0.1$. The legend indicates the length of the training time series in number of $\Delta t$ steps (i.e., $T/\Delta t$). For a comparison with a natural time scale of the KS system, we note that $10000$ $\Delta t$ time steps equals approximately $200$ Lyapunov times.}
\label{fig:vary_tl}
\end{figure}

We find that the amount of data used to train the reservoir computer can significantly affect the accuracy of the Lyapunov spectrum. The negative Lyapunov exponents are more sensitive than the positive exponents to errors due to insufficient training data. Figure~\ref{fig:vary_tl} demonstrates this result through a plot of the Lyapunov spectrum of the reservoir trained on varying lengths of data from Eq.~(\ref{eq:ks}) with parameters $L = 60$, $\lambda = 15$ and $\mu=0.1$. In this example we find that we need a training time series of greater than $20000$ time steps in order to obtain a reasonably accurate estimate of the negative Lyapunov exponents ($20000$ time steps equals about $400$ multiples of the Lyapunov time ($\Lambda_1^{-1}$) which can be considered to be a natural time scale of the KS system).

\section{\label{sec:conclusion}Discussion and Conclusion}
We conclude that a suitably trained reservoir computing system is capable of approximating the ergodic properties of the true system that it was trained on. Remarkably, as shown in Sec. \ref{sec:KS}, it is possible to use the trained reservoir to calculate a large number of positive and negative Lyapunov exponents of a high dimensional spatio-temporal chaotic system with good accuracy. In the case of the Lorenz equations, our method is successful in calculating the positive and zero Lyapunov exponents with good accuracy. The negative Lyapunov exponent of the true Lorenz system has a high magnitude, and our method is not as successful in accurately calculating the numerical value of this exponent, although it does successfully capture that its magnitude is substantially larger than that of the positive exponent. From a more general point of view, our paper suggests that the development of machine learning techniques for model-free analysis of measured data from chaotic systems may be a fruitful subject for further research.

\section{\label{sec:Ack}Acknowledgment}
This work was supported by grant W911NF1210101 from the Army Research Office. We wish to acknowledge useful conversations with Daniel Gauthier, Alexander Hartemink and Roger Brockett.


\begin{thebibliography}{30}%
\makeatletter
\providecommand \@ifxundefined [1]{%
 \@ifx{#1\undefined}
}%
\providecommand \@ifnum [1]{%
 \ifnum #1\expandafter \@firstoftwo
 \else \expandafter \@secondoftwo
 \fi
}%
\providecommand \@ifx [1]{%
 \ifx #1\expandafter \@firstoftwo
 \else \expandafter \@secondoftwo
 \fi
}%
\providecommand \natexlab [1]{#1}%
\providecommand \enquote  [1]{``#1''}%
\providecommand \bibnamefont  [1]{#1}%
\providecommand \bibfnamefont [1]{#1}%
\providecommand \citenamefont [1]{#1}%
\providecommand \href@noop [0]{\@secondoftwo}%
\providecommand \href [0]{\begingroup \@sanitize@url \@href}%
\providecommand \@href[1]{\@@startlink{#1}\@@href}%
\providecommand \@@href[1]{\endgroup#1\@@endlink}%
\providecommand \@sanitize@url [0]{\catcode `\\12\catcode `\$12\catcode
  `\&12\catcode `\#12\catcode `\^12\catcode `\_12\catcode `\%12\relax}%
\providecommand \@@startlink[1]{}%
\providecommand \@@endlink[0]{}%
\providecommand \url  [0]{\begingroup\@sanitize@url \@url }%
\providecommand \@url [1]{\endgroup\@href {#1}{\urlprefix }}%
\providecommand \urlprefix  [0]{URL }%
\providecommand \Eprint [0]{\href }%
\providecommand \doibase [0]{http://dx.doi.org/}%
\providecommand \selectlanguage [0]{\@gobble}%
\providecommand \bibinfo  [0]{\@secondoftwo}%
\providecommand \bibfield  [0]{\@secondoftwo}%
\providecommand \translation [1]{[#1]}%
\providecommand \BibitemOpen [0]{}%
\providecommand \bibitemStop [0]{}%
\providecommand \bibitemNoStop [0]{.\EOS\space}%
\providecommand \EOS [0]{\spacefactor3000\relax}%
\providecommand \BibitemShut  [1]{\csname bibitem#1\endcsname}%
\let\auto@bib@innerbib\@empty
\bibitem [{\citenamefont {Hinton}\ \emph {et~al.}(2012)\citenamefont {Hinton},
  \citenamefont {Deng}, \citenamefont {Yu}, \citenamefont {Dahl}, \citenamefont
  {Mohamed}, \citenamefont {Jaitly}, \citenamefont {Senior}, \citenamefont
  {Vanhoucke}, \citenamefont {Nguyen}, \citenamefont {Sainath} \emph
  {et~al.}}]{hintondeep}%
  \BibitemOpen
  \bibfield  {author} {\bibinfo {author} {\bibfnamefont {G.}~\bibnamefont
  {Hinton}}, \bibinfo {author} {\bibfnamefont {L.}~\bibnamefont {Deng}},
  \bibinfo {author} {\bibfnamefont {D.}~\bibnamefont {Yu}}, \bibinfo {author}
  {\bibfnamefont {G.~E.}\ \bibnamefont {Dahl}}, \bibinfo {author}
  {\bibfnamefont {A.-R.}\ \bibnamefont {Mohamed}}, \bibinfo {author}
  {\bibfnamefont {N.}~\bibnamefont {Jaitly}}, \bibinfo {author} {\bibfnamefont
  {A.}~\bibnamefont {Senior}}, \bibinfo {author} {\bibfnamefont
  {V.}~\bibnamefont {Vanhoucke}}, \bibinfo {author} {\bibfnamefont
  {P.}~\bibnamefont {Nguyen}}, \bibinfo {author} {\bibfnamefont {T.~N.}\
  \bibnamefont {Sainath}},  \emph {et~al.},\ }\href@noop {} {\bibfield
  {journal} {\bibinfo  {journal} {IEEE Signal Processing Magazine}\ }\textbf
  {\bibinfo {volume} {29}},\ \bibinfo {pages} {82} (\bibinfo {year}
  {2012})}\BibitemShut {NoStop}%
\bibitem [{\citenamefont {Goodfellow}\ \emph {et~al.}(2016)\citenamefont
  {Goodfellow}, \citenamefont {Bengio},\ and\ \citenamefont
  {Courville}}]{goodfellow}%
  \BibitemOpen
  \bibfield  {author} {\bibinfo {author} {\bibfnamefont {I.}~\bibnamefont
  {Goodfellow}}, \bibinfo {author} {\bibfnamefont {Y.}~\bibnamefont {Bengio}},
  \ and\ \bibinfo {author} {\bibfnamefont {A.}~\bibnamefont {Courville}},\
  }\href@noop {} {\emph {\bibinfo {title} {Deep Learning}}}\ (\bibinfo
  {publisher} {MIT Press},\ \bibinfo {year} {2016})\BibitemShut {NoStop}%
\bibitem [{\citenamefont {Silver}\ \emph {et~al.}(2016)\citenamefont {Silver},
  \citenamefont {Huang}, \citenamefont {Maddison}, \citenamefont {Guez},
  \citenamefont {Sifre}, \citenamefont {Van Den~Driessche}, \citenamefont
  {Schrittwieser}, \citenamefont {Antonoglou}, \citenamefont {Panneershelvam},
  \citenamefont {Lanctot} \emph {et~al.}}]{silvergo}%
  \BibitemOpen
  \bibfield  {author} {\bibinfo {author} {\bibfnamefont {D.}~\bibnamefont
  {Silver}}, \bibinfo {author} {\bibfnamefont {A.}~\bibnamefont {Huang}},
  \bibinfo {author} {\bibfnamefont {C.~J.}\ \bibnamefont {Maddison}}, \bibinfo
  {author} {\bibfnamefont {A.}~\bibnamefont {Guez}}, \bibinfo {author}
  {\bibfnamefont {L.}~\bibnamefont {Sifre}}, \bibinfo {author} {\bibfnamefont
  {G.}~\bibnamefont {Van Den~Driessche}}, \bibinfo {author} {\bibfnamefont
  {J.}~\bibnamefont {Schrittwieser}}, \bibinfo {author} {\bibfnamefont
  {I.}~\bibnamefont {Antonoglou}}, \bibinfo {author} {\bibfnamefont
  {V.}~\bibnamefont {Panneershelvam}}, \bibinfo {author} {\bibfnamefont
  {M.}~\bibnamefont {Lanctot}},  \emph {et~al.},\ }\href@noop {} {\bibfield
  {journal} {\bibinfo  {journal} {Nature}\ }\textbf {\bibinfo {volume} {529}},\
  \bibinfo {pages} {484} (\bibinfo {year} {2016})}\BibitemShut {NoStop}%
\bibitem [{\citenamefont {Lukosevivcius}\ and\ \citenamefont
  {Jaeger}(2009)}]{lukoreservoir}%
  \BibitemOpen
  \bibfield  {author} {\bibinfo {author} {\bibfnamefont {M.}~\bibnamefont
  {Lukosevivcius}}\ and\ \bibinfo {author} {\bibfnamefont {H.}~\bibnamefont
  {Jaeger}},\ }\href@noop {} {\bibfield  {journal} {\bibinfo  {journal}
  {Computer Science Review}\ }\textbf {\bibinfo {volume} {3}},\ \bibinfo
  {pages} {127} (\bibinfo {year} {2009})}\BibitemShut {NoStop}%
\bibitem [{\citenamefont {Kantz}\ and\ \citenamefont
  {Schreiber}(2004)}]{kantzNonlinear}%
  \BibitemOpen
  \bibfield  {author} {\bibinfo {author} {\bibfnamefont {H.}~\bibnamefont
  {Kantz}}\ and\ \bibinfo {author} {\bibfnamefont {T.}~\bibnamefont
  {Schreiber}},\ }\href@noop {} {\emph {\bibinfo {title} {Nonlinear time series
  analysis}}},\ Vol.~\bibinfo {volume} {7}\ (\bibinfo  {publisher} {Cambridge
  University Press},\ \bibinfo {year} {2004})\BibitemShut {NoStop}%
\bibitem [{\citenamefont {Ott}\ \emph {et~al.}(1994)\citenamefont {Ott},
  \citenamefont {Sauer},\ and\ \citenamefont {Yorke}}]{ottCoping}%
  \BibitemOpen
  \bibfield  {author} {\bibinfo {author} {\bibfnamefont {E.}~\bibnamefont
  {Ott}}, \bibinfo {author} {\bibfnamefont {T.}~\bibnamefont {Sauer}}, \ and\
  \bibinfo {author} {\bibfnamefont {J.~A.}\ \bibnamefont {Yorke}},\ }\href@noop
  {} {\bibfield  {journal} {\bibinfo  {journal} {Wiley Series in Nonlinear
  Science, New York: John Wiley}\ } (\bibinfo {year} {1994})}\BibitemShut
  {NoStop}%
\bibitem [{\citenamefont {Abarbanel}(2012)}]{abarbanelAnalysis}%
  \BibitemOpen
  \bibfield  {author} {\bibinfo {author} {\bibfnamefont {H.}~\bibnamefont
  {Abarbanel}},\ }\href@noop {} {\emph {\bibinfo {title} {Analysis of observed
  chaotic data}}}\ (\bibinfo  {publisher} {Springer Science \& Business
  Media},\ \bibinfo {year} {2012})\BibitemShut {NoStop}%
\bibitem [{\citenamefont {Takens}(1981)}]{takens}%
  \BibitemOpen
  \bibfield  {author} {\bibinfo {author} {\bibfnamefont {F.}~\bibnamefont
  {Takens}},\ }\href@noop {} {\bibfield  {journal} {\bibinfo  {journal} {by DA
  Rand and L.-S. Young Springer, Berlin}\ }\textbf {\bibinfo {volume} {898}},\
  \bibinfo {pages} {366} (\bibinfo {year} {1981})}\BibitemShut {NoStop}%
\bibitem [{\citenamefont {Sauer}\ \emph {et~al.}(1991)\citenamefont {Sauer},
  \citenamefont {Yorke},\ and\ \citenamefont {Casdagli}}]{sauerEmbedology}%
  \BibitemOpen
  \bibfield  {author} {\bibinfo {author} {\bibfnamefont {T.}~\bibnamefont
  {Sauer}}, \bibinfo {author} {\bibfnamefont {J.~A.}\ \bibnamefont {Yorke}}, \
  and\ \bibinfo {author} {\bibfnamefont {M.}~\bibnamefont {Casdagli}},\
  }\href@noop {} {\bibfield  {journal} {\bibinfo  {journal} {Journal of
  Statistical Physics}\ }\textbf {\bibinfo {volume} {65}},\ \bibinfo {pages}
  {579} (\bibinfo {year} {1991})}\BibitemShut {NoStop}%
\bibitem [{\citenamefont {Broomhead}\ and\ \citenamefont
  {King}(1986)}]{broomheadExtracting}%
  \BibitemOpen
  \bibfield  {author} {\bibinfo {author} {\bibfnamefont {D.~S.}\ \bibnamefont
  {Broomhead}}\ and\ \bibinfo {author} {\bibfnamefont {G.~P.}\ \bibnamefont
  {King}},\ }\href@noop {} {\bibfield  {journal} {\bibinfo  {journal} {Physica
  D: Nonlinear Phenomena}\ }\textbf {\bibinfo {volume} {20}},\ \bibinfo {pages}
  {217} (\bibinfo {year} {1986})}\BibitemShut {NoStop}%
\bibitem [{\citenamefont {Brandstater}\ and\ \citenamefont
  {Swinney}(1987)}]{brandstatedStrange}%
  \BibitemOpen
  \bibfield  {author} {\bibinfo {author} {\bibfnamefont {A.}~\bibnamefont
  {Brandstater}}\ and\ \bibinfo {author} {\bibfnamefont {H.~L.}\ \bibnamefont
  {Swinney}},\ }\href@noop {} {\bibfield  {journal} {\bibinfo  {journal}
  {Physical Review A}\ }\textbf {\bibinfo {volume} {35}},\ \bibinfo {pages}
  {2207} (\bibinfo {year} {1987})}\BibitemShut {NoStop}%
\bibitem [{\citenamefont {Eckmann}\ \emph {et~al.}(1986)\citenamefont
  {Eckmann}, \citenamefont {Kamphorst}, \citenamefont {Ruelle},\ and\
  \citenamefont {Ciliberto}}]{eckmannLiapunov}%
  \BibitemOpen
  \bibfield  {author} {\bibinfo {author} {\bibfnamefont {J.-P.}\ \bibnamefont
  {Eckmann}}, \bibinfo {author} {\bibfnamefont {S.~O.}\ \bibnamefont
  {Kamphorst}}, \bibinfo {author} {\bibfnamefont {D.}~\bibnamefont {Ruelle}}, \
  and\ \bibinfo {author} {\bibfnamefont {S.}~\bibnamefont {Ciliberto}},\
  }\href@noop {} {\bibfield  {journal} {\bibinfo  {journal} {Physical Review
  A}\ }\textbf {\bibinfo {volume} {34}},\ \bibinfo {pages} {4971} (\bibinfo
  {year} {1986})}\BibitemShut {NoStop}%
\bibitem [{\citenamefont {Petrov}\ \emph {et~al.}(1993)\citenamefont {Petrov},
  \citenamefont {Gaspar}, \citenamefont {Masere},\ and\ \citenamefont
  {Showalter}}]{petrovControlling}%
  \BibitemOpen
  \bibfield  {author} {\bibinfo {author} {\bibfnamefont {V.}~\bibnamefont
  {Petrov}}, \bibinfo {author} {\bibfnamefont {V.}~\bibnamefont {Gaspar}},
  \bibinfo {author} {\bibfnamefont {J.}~\bibnamefont {Masere}}, \ and\ \bibinfo
  {author} {\bibfnamefont {K.}~\bibnamefont {Showalter}},\ }\href@noop {}
  {\bibfield  {journal} {\bibinfo  {journal} {Nature}\ }\textbf {\bibinfo
  {volume} {361}},\ \bibinfo {pages} {240} (\bibinfo {year}
  {1993})}\BibitemShut {NoStop}%
\bibitem [{\citenamefont {Lorenz}(1963)}]{lorenzDeterministic}%
  \BibitemOpen
  \bibfield  {author} {\bibinfo {author} {\bibfnamefont {E.~N.}\ \bibnamefont
  {Lorenz}},\ }\href@noop {} {\bibfield  {journal} {\bibinfo  {journal}
  {Journal of the Atmospheric Sciences}\ }\textbf {\bibinfo {volume} {20}},\
  \bibinfo {pages} {130} (\bibinfo {year} {1963})}\BibitemShut {NoStop}%
\bibitem [{\citenamefont {Cohen}\ \emph {et~al.}(1976)\citenamefont {Cohen},
  \citenamefont {Krommes}, \citenamefont {Tang},\ and\ \citenamefont
  {Rosenbluth}}]{cohenNonlinear}%
  \BibitemOpen
  \bibfield  {author} {\bibinfo {author} {\bibfnamefont {B.~I.}\ \bibnamefont
  {Cohen}}, \bibinfo {author} {\bibfnamefont {J.}~\bibnamefont {Krommes}},
  \bibinfo {author} {\bibfnamefont {W.}~\bibnamefont {Tang}}, \ and\ \bibinfo
  {author} {\bibfnamefont {M.}~\bibnamefont {Rosenbluth}},\ }\href@noop {}
  {\bibfield  {journal} {\bibinfo  {journal} {Nuclear Fusion}\ }\textbf
  {\bibinfo {volume} {16}},\ \bibinfo {pages} {971} (\bibinfo {year}
  {1976})}\BibitemShut {NoStop}%
\bibitem [{\citenamefont {Kuramoto}\ and\ \citenamefont
  {Tsuzuki}(1976)}]{kuramotoPersistent}%
  \BibitemOpen
  \bibfield  {author} {\bibinfo {author} {\bibfnamefont {Y.}~\bibnamefont
  {Kuramoto}}\ and\ \bibinfo {author} {\bibfnamefont {T.}~\bibnamefont
  {Tsuzuki}},\ }\href@noop {} {\bibfield  {journal} {\bibinfo  {journal}
  {Progress of Theoretical Physics}\ }\textbf {\bibinfo {volume} {55}},\
  \bibinfo {pages} {356} (\bibinfo {year} {1976})}\BibitemShut {NoStop}%
\bibitem [{\citenamefont {Sivashinsky}(1982)}]{sivashinskyLarge}%
  \BibitemOpen
  \bibfield  {author} {\bibinfo {author} {\bibfnamefont {G.}~\bibnamefont
  {Sivashinsky}},\ }\href@noop {} {\bibfield  {journal} {\bibinfo  {journal}
  {Physica D: Nonlinear Phenomena}\ }\textbf {\bibinfo {volume} {4}},\ \bibinfo
  {pages} {227} (\bibinfo {year} {1982})}\BibitemShut {NoStop}%
\bibitem [{\citenamefont {Cross}\ and\ \citenamefont
  {Hohenberg}(1993)}]{crossPattern}%
  \BibitemOpen
  \bibfield  {author} {\bibinfo {author} {\bibfnamefont {M.~C.}\ \bibnamefont
  {Cross}}\ and\ \bibinfo {author} {\bibfnamefont {P.~C.}\ \bibnamefont
  {Hohenberg}},\ }\href@noop {} {\bibfield  {journal} {\bibinfo  {journal}
  {Reviews of Modern Physics}\ }\textbf {\bibinfo {volume} {65}},\ \bibinfo
  {pages} {851} (\bibinfo {year} {1993})}\BibitemShut {NoStop}%
\bibitem [{\citenamefont {Livi}\ \emph {et~al.}(1986)\citenamefont {Livi},
  \citenamefont {Politi},\ and\ \citenamefont {Ruffo}}]{liviDistribution}%
  \BibitemOpen
  \bibfield  {author} {\bibinfo {author} {\bibfnamefont {R.}~\bibnamefont
  {Livi}}, \bibinfo {author} {\bibfnamefont {A.}~\bibnamefont {Politi}}, \ and\
  \bibinfo {author} {\bibfnamefont {S.}~\bibnamefont {Ruffo}},\ }\href@noop {}
  {\bibfield  {journal} {\bibinfo  {journal} {Journal of Physics A:
  Mathematical and General}\ }\textbf {\bibinfo {volume} {19}},\ \bibinfo
  {pages} {2033} (\bibinfo {year} {1986})}\BibitemShut {NoStop}%
\bibitem [{\citenamefont {Egolf}\ and\ \citenamefont
  {Greenside}(1994)}]{egolfRelation}%
  \BibitemOpen
  \bibfield  {author} {\bibinfo {author} {\bibfnamefont {D.~A.}\ \bibnamefont
  {Egolf}}\ and\ \bibinfo {author} {\bibfnamefont {H.~S.}\ \bibnamefont
  {Greenside}},\ }\href@noop {} {\bibfield  {journal} {\bibinfo  {journal}
  {Nature}\ }\textbf {\bibinfo {volume} {369}},\ \bibinfo {pages} {129}
  (\bibinfo {year} {1994})}\BibitemShut {NoStop}%
\bibitem [{\citenamefont {Pikovsky}\ and\ \citenamefont
  {Politi}(1998)}]{pikovskyDynamic}%
  \BibitemOpen
  \bibfield  {author} {\bibinfo {author} {\bibfnamefont {A.}~\bibnamefont
  {Pikovsky}}\ and\ \bibinfo {author} {\bibfnamefont {A.}~\bibnamefont
  {Politi}},\ }\href@noop {} {\bibfield  {journal} {\bibinfo  {journal}
  {Nonlinearity}\ }\textbf {\bibinfo {volume} {11}},\ \bibinfo {pages} {1049}
  (\bibinfo {year} {1998})}\BibitemShut {NoStop}%
\bibitem [{\citenamefont {Kantz}\ \emph {et~al.}(2013)\citenamefont {Kantz},
  \citenamefont {Radons},\ and\ \citenamefont {Yang}}]{kantz2013problem}%
  \BibitemOpen
  \bibfield  {author} {\bibinfo {author} {\bibfnamefont {H.}~\bibnamefont
  {Kantz}}, \bibinfo {author} {\bibfnamefont {G.}~\bibnamefont {Radons}}, \
  and\ \bibinfo {author} {\bibfnamefont {H.}~\bibnamefont {Yang}},\ }\href@noop
  {} {\bibfield  {journal} {\bibinfo  {journal} {Journal of Physics A:
  Mathematical and Theoretical}\ }\textbf {\bibinfo {volume} {46}},\ \bibinfo
  {pages} {254009} (\bibinfo {year} {2013})}\BibitemShut {NoStop}%
\bibitem [{\citenamefont {Sauer}\ \emph {et~al.}(1998)\citenamefont {Sauer},
  \citenamefont {Tempkin},\ and\ \citenamefont {Yorke}}]{sauer1998spurious}%
  \BibitemOpen
  \bibfield  {author} {\bibinfo {author} {\bibfnamefont {T.~D.}\ \bibnamefont
  {Sauer}}, \bibinfo {author} {\bibfnamefont {J.~A.}\ \bibnamefont {Tempkin}},
  \ and\ \bibinfo {author} {\bibfnamefont {J.~A.}\ \bibnamefont {Yorke}},\
  }\href@noop {} {\bibfield  {journal} {\bibinfo  {journal} {Physical Review
  Letters}\ }\textbf {\bibinfo {volume} {81}},\ \bibinfo {pages} {4341}
  (\bibinfo {year} {1998})}\BibitemShut {NoStop}%
\bibitem [{\citenamefont {Jaeger}\ and\ \citenamefont
  {Haas}(2004)}]{jaegerHarnessing}%
  \BibitemOpen
  \bibfield  {author} {\bibinfo {author} {\bibfnamefont {H.}~\bibnamefont
  {Jaeger}}\ and\ \bibinfo {author} {\bibfnamefont {H.}~\bibnamefont {Haas}},\
  }\href@noop {} {\bibfield  {journal} {\bibinfo  {journal} {Science}\ }\textbf
  {\bibinfo {volume} {304}},\ \bibinfo {pages} {78} (\bibinfo {year}
  {2004})}\BibitemShut {NoStop}%
\bibitem [{\citenamefont {Lu}\ \emph {et~al.}(2017)\citenamefont {Lu},
  \citenamefont {Pathak}, \citenamefont {Hunt}, \citenamefont {Girvan},
  \citenamefont {Brockett},\ and\ \citenamefont {Ott}}]{lu2017reservoir}%
  \BibitemOpen
  \bibfield  {author} {\bibinfo {author} {\bibfnamefont {Z.}~\bibnamefont
  {Lu}}, \bibinfo {author} {\bibfnamefont {J.}~\bibnamefont {Pathak}}, \bibinfo
  {author} {\bibfnamefont {B.}~\bibnamefont {Hunt}}, \bibinfo {author}
  {\bibfnamefont {M.}~\bibnamefont {Girvan}}, \bibinfo {author} {\bibfnamefont
  {R.}~\bibnamefont {Brockett}}, \ and\ \bibinfo {author} {\bibfnamefont
  {E.}~\bibnamefont {Ott}},\ }\href@noop {} {\bibfield  {journal} {\bibinfo
  {journal} {Chaos: An Interdisciplinary Journal of Nonlinear Science}\
  }\textbf {\bibinfo {volume} {27}},\ \bibinfo {pages} {041102} (\bibinfo
  {year} {2017})}\BibitemShut {NoStop}%
\bibitem [{\citenamefont {Jaeger}(2001)}]{jaegerEcho}%
  \BibitemOpen
  \bibfield  {author} {\bibinfo {author} {\bibfnamefont {H.}~\bibnamefont
  {Jaeger}},\ }\href@noop {} {\bibfield  {journal} {\bibinfo  {journal} {Bonn,
  Germany: German National Research Center for Information Technology GMD
  Technical Report}\ }\textbf {\bibinfo {volume} {148}},\ \bibinfo {pages} {13}
  (\bibinfo {year} {2001})}\BibitemShut {NoStop}%
\bibitem [{\citenamefont {Maass}\ \emph {et~al.}(2002)\citenamefont {Maass},
  \citenamefont {Natschlager},\ and\ \citenamefont {Markram}}]{maassReal}%
  \BibitemOpen
  \bibfield  {author} {\bibinfo {author} {\bibfnamefont {W.}~\bibnamefont
  {Maass}}, \bibinfo {author} {\bibfnamefont {T.}~\bibnamefont {Natschlager}},
  \ and\ \bibinfo {author} {\bibfnamefont {H.}~\bibnamefont {Markram}},\
  }\href@noop {} {\bibfield  {journal} {\bibinfo  {journal} {Neural
  Computation}\ }\textbf {\bibinfo {volume} {14}},\ \bibinfo {pages} {2531}
  (\bibinfo {year} {2002})}\BibitemShut {NoStop}%
\bibitem [{\citenamefont {Tikhonov}\ \emph {et~al.}(1977)\citenamefont
  {Tikhonov}, \citenamefont {Arsenin},\ and\ \citenamefont
  {John}}]{tikhonov1977solutions}%
  \BibitemOpen
  \bibfield  {author} {\bibinfo {author} {\bibfnamefont {N.}~\bibnamefont
  {Tikhonov}, \bibfnamefont {Andre{\u\i}}}, \bibinfo {author} {\bibfnamefont
  {V.~I.}\ \bibnamefont {Arsenin}}, \ and\ \bibinfo {author} {\bibfnamefont
  {F.}~\bibnamefont {John}},\ }\href@noop {} {\emph {\bibinfo {title}
  {Solutions of ill-posed problems}}},\ Vol.~\bibinfo {volume} {14}\ (\bibinfo
  {publisher} {Winston Washington, DC},\ \bibinfo {year} {1977})\BibitemShut
  {NoStop}%
\bibitem [{\citenamefont {Yan}\ and\ \citenamefont {Su}(2009)}]{yan2009linear}%
  \BibitemOpen
  \bibfield  {author} {\bibinfo {author} {\bibfnamefont {X.}~\bibnamefont
  {Yan}}\ and\ \bibinfo {author} {\bibfnamefont {X.}~\bibnamefont {Su}},\
  }\href@noop {} {\emph {\bibinfo {title} {Linear regression analysis: theory
  and computing}}}\ (\bibinfo  {publisher} {World Scientific},\ \bibinfo {year}
  {2009})\BibitemShut {NoStop}%
\bibitem [{\citenamefont {Kaplan}\ and\ \citenamefont
  {Yorke}(1979)}]{kaplan1979chaotic}%
  \BibitemOpen
  \bibfield  {author} {\bibinfo {author} {\bibfnamefont {J.~L.}\ \bibnamefont
  {Kaplan}}\ and\ \bibinfo {author} {\bibfnamefont {J.~A.}\ \bibnamefont
  {Yorke}},\ }in\ \href@noop {} {\emph {\bibinfo {booktitle} {Functional
  Differential Equations and Approximation of Fixed Points}}}\ (\bibinfo
  {publisher} {Springer},\ \bibinfo {year} {1979})\ pp.\ \bibinfo {pages}
  {204--227}\BibitemShut {NoStop}%
\end{thebibliography}
\end{document}